\documentclass[twocolumn,showpacs,preprintnumbers,amsmath,amssymb,aps,floatfix,superscriptaddress]{revtex4}
\usepackage{gensymb,amsmath,amssymb}
\usepackage{graphicx}
\usepackage{dcolumn}
\usepackage{bm}
\usepackage{units}
\usepackage[usenames]{color}
\usepackage{float}

\begin{document}

\title{Strain engineering of topological properties in lead-salt semiconductors}

\author{Paolo Barone}
\affiliation{Consiglio Nazionale delle Ricerche (CNR-SPIN), I-67100 L'Aquila, Italy}

\author{Domenico Di Sante}
\affiliation{Consiglio Nazionale delle Ricerche (CNR-SPIN), I-67100 L'Aquila, Italy}
\affiliation{Department of Physical and Chemical Sciences, University of L'Aquila, Via Vetoio 10, I-67010 L'Aquila, Italy}

\author{Silvia Picozzi}
\affiliation{Consiglio Nazionale delle Ricerche (CNR-SPIN), I-67100 L'Aquila, Italy}


\begin{abstract}
Rock-salt chalcogenide SnTe represents the simplest realization of a topological insulator where a crystal symmetry
allows for the appearence of topologically protected metallic states with an even number of Dirac cones on
high-symmetry crystal surfaces. Related rock-salt lead
chalcogenides have been predicted as well to undergo a phase-transition to a topological crystalline insulating phase after
band inversion induced by alloying and pressure. Here we theoretically predict that strain, as realized in thin films grown on
(001) substrates, may induce such topological phase-transitions. Furthermore, relevant
topological properties of the surface states, such as the location of the Dirac cones on
the surface Brillouin zone or the decay length of edge states, appear to be tunable with strain,
with potential implications for technological devices benefiting from those additional degrees of freedom.
\end{abstract}

\maketitle


Topological band-insulators are new states of matter where the electronic structures display non-trivial
topological features leading to the appearence of symmetry-protected metallic edge states which
feature graphene-like Dirac cones. \cite{Hasan_RMP,Qi_RMP,Moore_nature,Fu_PRL}
As a result of their non-trivial topological invariants
(based on the wave-function ``wrapping'' of the Brillouin zone), these surface states are
exceptionally robust against perturbations. Moreover, the spin-orbit coupling (SOC), which ultimately lies at the
origin of the phenomenon, makes the surface states to show a single spin-state for each momentum,
paving the way to spin-polarized currents of interest in spintronics.\cite{Hasan_RMP,Qi_RMP,Moore_nature,Fu_PRL}
After the discovery of the so-called $Z_2$ topological insulators, where the unusual metallic states are protected by
time-reversal symmetry, it has been realized that other fundamental symmetries may allow for the existence of
topologically-protected states of matter\cite{Fu_TCI,Slager}. 
In particular, it has been shown that the narrow-gap
IV-VI semiconductor SnTe is a topological crystalline insulator (TCI)
in its face-centered-cubic ($fcc$) structure, where a mirror symmetry is responsible for the appearence of an
even number of Dirac cones on (001), (011) and (111) surfaces\cite{Hsieh_ncomm,Tanaka_nature}.
With the only exception of (111) surfaces, these Dirac cones are situated, in contrast to $Z_2$ topological insulators,
off the time-reversal-invariant momenta
along a high-symmetry line of the surface Brillouin zone contained in a mirror
plane of the $fcc$ crystal structure.
As reported recently, this unique feature of SnTe arises from the hybridization of two distinct Dirac cones projected on
the (001) and (011) surfaces\cite{Tanaka_nature,Fu_111,Tanaka_prb}. 

Even though closely related lead-salt semiconductors PbS, PbSe and PbTe display the same symmetry properties as SnTe, their
electronic structures appear to be topologically trivial at equilibrium volume. Metallic surface
states in these compounds do not show up due to the absence of a bulk band inversion, that is generally
believed to be caused by SOC\cite{Hsieh_ncomm}.
So far, the possibility to induce topological transitions in these systems has been explored by alloying lead
chalcogenides PbTe and PbSe with Sn, thus triggering the required band inversion while keeping the same
crystal structure\cite{Tanaka_prb,Xu_ncomm,Dziawa_nmat,Wojek}.
Interestingly, the location of the Dirac cones on the (001) surfaces of Pb$_{1-x}$Sn$_x$Te has been found to vary
in $\bm k$ space when changing the composition ratio $x$, providing an additional degree of freedom
with potential applications for spintronic devices\cite{Tanaka_prb}.
On the other hand, it has been recently suggested that the band inversion arises from the interplay of SOC
and an ``asymmetric'' hybridization between cation and anion $sp$ orbitals\cite{nostro}; as such, a topological phase-transition
could be naturally attained by applying external pressure or strain, which would directly affect the aforementioned hybridization,
as long as the mirror symmetries are preserved. Specifically, all lead chalcogenides have been predicted to become
TCIs upon applied pressures in the $\unit{GPa}$ range\cite{nostro}. 
The effect of strain on the topological properties of this class of narrow-band semiconductors,
however, has not been addressed in details yet, even though a strain-induced TCI transition has been recently predicted in
PbTe\cite{huang_matexp}, whereas strain engineering of $Z_2$ topological order has been
already considered in several materials including
ternary Heusler, zinc-blende and cubic semiconducting compounds\cite{Felser_zhang,winterfield,feng}.

Films of lead-salt chalcogenides have been long fabricated using a variety of techniques, ranging from liquid phase epitaxy
to molecular beam epitaxy, and several
substrates have been used\cite{superlattices,MBE,Mukherjee}. 
Due to their rock-salt structures, lead-salt compounds have (001) natural cleavage planes and tend to grow in the (001) growth
orientation, although the (111) growth orientation has also been reported.
 Alkali-halide substrates such as NaCl, NaBr, NaI, KCl, KBr, KI, LiF have been widely used
for IV-VI-compound (001)-oriented epitaxial thin films\cite{superlattices}, since they mainly have the same rock-salt
structure and comparable lattice
constants; other substrates such as (111)Si or BaF$_2$ have been used for the growth of (111)-oriented films\cite{MBE}.

In this work we theoretically demonstrate, by means of {\it ab initio}-based tight-binding (TB) and accurate
density-functional theory (DFT) calculations, that all lead-salt chalcogenides may be turned into topological insulators at
reasonable values of strain when grown on (001)-oriented substrates. 
The $\bm k$-space location of the Dirac cones appear to be dependent on the strain, thus suggesting an alternative path 
for the tunability of topological properties in TCIs, as opposed to alloying or pressure,
and envisaging novel spintronic applications based e.g. on piezoelectric substrates. The critical thickness at which
topological edge states appear is found to be of the order of $\sim \unit[100]{nm}$ close to the strain-induced topological
transition. However, as the bulk gap increases upon applying larger strain, the decay length of surface states is largely
reduced, suggesting that topological thin films may be engineered by a proper choice of the substrate. On the other hand,
in films epitaxially grown on (111)
substrates an insulator to semi-metal transition, triggered by a rhombohedral distortion, is predicted to occur before
any band inversion --- and hence any topological transition --- can appear. 

{\bf Methods.}
The systematic analysis of strain effects in the rock-salt IV-VI class of compounds has been performed by resorting
to an {\it ab initio}-based relativistic parametrization of a nearest-neighbor ($nn$) TB model including 
$s$, $p$ and $d$ states\cite{Lent}. Structural deformations as induced by strain have been taken into account
assuming an ideal thin strained overlayer grown over a substrate which determines the
lattice constant parallel to the interface plane. Within these assumptions, interfacial strain and perpendicular strained
lattice constant are given by $\delta=\nicefrac{a}{a_0}-1$ and $c=c_0\left[1-D^i\delta\right]$, where $a_0, c_0$ are
the equilibrium lattice constants and the constants $D^i$ have been estimated from
the experimental elastic constants $c_{11}, c_{12}$ and $c_{44}$\cite{elastic1,elastic2} as\cite{VandeWalle}:
\begin{eqnarray}
D^{001}&=&2\frac{c_{12}}{c_{11}},\\
D^{111}&=&2\frac{c_{11}+2c_{12}-2c_{44}}{c_{11}+2c_{12}+4c_{44}}.
\end{eqnarray}
 All TB parameters have been
accordingly rescaled following Harrison's rules\cite{Harrison}, $t_{nm}(\delta)=\nicefrac{t^0_{nm}}{r(\delta)^\alpha}$,
where $r(\delta)$ are the ion-ion distances modified by the applied strain, $n,m= s, p, d$ label the orbital
states and $\alpha=2, 3.5, 5$
are the scaling exponents for $\{s$-$s,s$-$p,p$-$p\}$, $\{s$-$d,p$-$d\}$ and $d$-$d$ hopping
interactions, respectively.

In order to benchmark the reliability of our TB calculations, 
we performed accurate DFT calculations with HSE
hybrid functional\cite{hybrid}, as implemented in VASP \cite{vasp1,vasp2}, for selected values of the strain.
Hybrid functionals improve signicantly with respect to the
local (LDA) and semi-local (GGA) approximations, especially for narrow
band-gap semiconductors and lead chalcogenides\cite{Cardona,Kresse}.
Total-energy calculations were performed to optimize the out-of-plane lattice parameter,
structure while progressively changing in-plane lattice constants in order to mimic strain conditions imposed by a
well-defined substrate. The energy cutoff for the
plane-wave expansion was 600 eV and an $8\times8\times8$ Monkhorst-
Pack k-point grid was used. Calculations with the HSE hybrid
functional are computationally very demanding and, thus, were used for bulk states only; however,  realistic
and accurate surface states have been calculated by means of a Wannier-functions parametrization of DFT band structure,
as derived via a Maximally-localized Wannier function algorithm\cite{wannier90}.

{\bf Results.}
Lead chalcogenides Pb$A$, with $A=$S, Se, Te, have a simple rock-salt structure with the fundamental band gap located at
four equivalent $L$ points in the rhombohedral setting. At their equilibrium volumes, they are not TCIs,
the ordering of conduction and valence bands not being inverted (as opposed to SnTe).
However, they are well known for their peculiar electronic structures, the small band-gap arising from
a strong level repulsion at the $L$ points of the $fcc$ Brillouin zone due to the presence of occupied Pb-6$s$ band
below the top of the valence band\cite{Wei,Cardona,Kresse}.
The band ordering close to the Fermi level is then determined by the interplay of $sp$ hybridizations between cation $s$
($p$) states with anion $p$ ($s$) states, that has been predicted to be strongly asymmetric in lead-salt chalcogenides and
SnTe, whereas a sizeable SOC eventually causes the opening of the gap\cite{nostro,huang_matexp}.
In the trivial insulating phase, which is smoothly connected to the atomic limit of infinitely far ions,
the valence band maximum (VBM) is expected to have strong Pb-$s$ and anion-$p$ character,
while the conductance band minimum (CBM) displays strong Pb-$p$ and anion-$s$ character. For the TCI SnTe, the ordering is reversed mainly because of a stronger $sp$ hybridization (also due to its smaller
equilibrium volume as opposed to PbTe), rather than by SOC alone (which is indeed smaller in Sn than in Pb).

The presence of mirror planes in the $fcc$ structure combined with
band-inversion at $L$ points implies a non-zero mirror Chern number which, in turn, dictates the existence of
spin-polarized surface states with opposite mirror eigenvalues appearing on (001), (011) and (111) surfaces which
preserve the relevant mirror-symmetry operations\cite{Hsieh_ncomm, Fu_111,Safaei}.
When considering strain in (001) and (111) oriented films, it is more
convenient to discuss their crystal and electronic structures in a tetragonal and hexagonal setting respectively,
with the fundamental
gap located at four equivalent $R$ points or $A$ and three equivalent $L$ points, respectively. All these
high-symmetry points are still related by the relevant mirror simmetries giving rise to nontrivial mirror Chern numbers. As a consequence, if a band
inversion is induced by biaxial strain in the (001) plane, a TCI is realized. On the other hand, biaxial strain
in the (111) plane may well cause a rhombohedral distortion, and in principle a band inversion may appear in
an even/odd number high-symmetry point ($A$ {\sl and/or} $L$), thus meaning the onset of a TCI/$Z_2$ topological
order.

{\it (001)-oriented systems.}
The strain deformation mainly affects the
anion-cation distances, and consequently their hopping interactions; specifically,
a compressive in-plane strain reduces the distances in the $(001)$ plane while increasing the distances
along the perpendicular axis
(a tensile strain doing the opposite). Furthermore, a significant volume effect is predicted as
$V(\delta)=V_0\,(1+\delta)^2\,(1-2D^{001}\delta)$, implying a positive/negative pressure effect
for compressive/tensile strain. We verified the accuracy of the
predicted structural deformations by comparing with {\it ab initio} calculations, which systematically predict 
larger volumes in PbTe and PbSe (1.4\%
and 0.5\% at $\delta=-4\%$, respectively), while a slightly smaller volume is found in PbS (-0.4\% at $\delta=-4\%$). 

\begin{figure}
\includegraphics[width=\columnwidth]{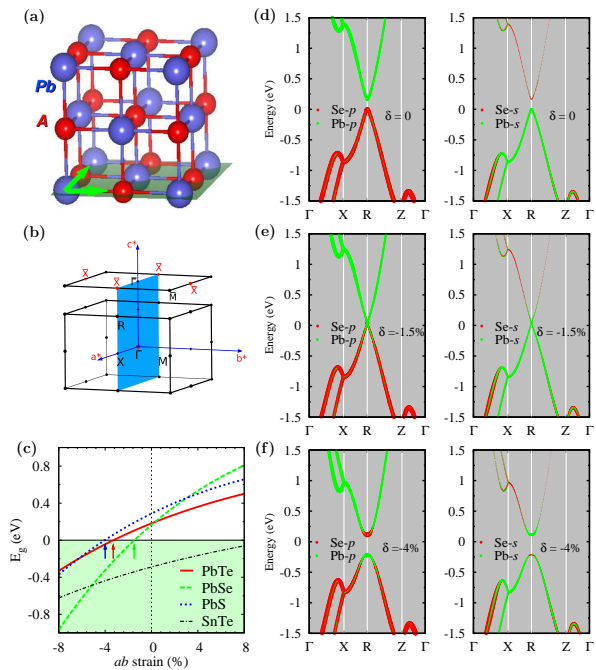}
\caption{(a)
Chosen unit cell for (001)-oriented films
with highlighted in-plane biaxial strain directions and (b) corresponding Brillouin zone. (c) Evolution of band gaps in lead-salt semiconductors and SnTe.
(d)-(f) Band structure evolution
of PbSe as a function of compressive
strain with projected Pb and Se $s,p$ states.  }\label{fig1}
\end{figure}

The combination of these structural deformations, directly acting upon the asymmetric $sp$
hybridization\cite{nostro}, causes a closure of the gap
followed by a reopening with an inversion of conductance and valence bands upon compression. Conversely,
upon tensile strain the band gap increases for
Pb$A$ non-inverted band structures mainly because of the ``negative
pressure'' effect (see Fig. \ref{fig1}).
The estimated critical strains from TB parametrizations are $\delta_c=-3.3\%, -1.5\%, -4.1\%$ for
$A=$Te, Se, S respectively. Our DFT calculations predict slightly larger critical strains for PbTe ($\sim -4\%$) and
PbSe ($\sim-2\%$), while at $-4\%$ the band inversion in PbS has already occurred.
The projected
$s$ and $p$ characters of the bands, as shown in Fig. \ref{fig1} for PbSe, further support the role of the asymmetric
hybridization mechanism in causing the
band-inversion phenomenon. 
(001)-strained lead-salt chalcogenides stay insulating at least up to a compressive strain of the order of $-10\%$.

\begin{figure}[b]
\includegraphics[width=\columnwidth]{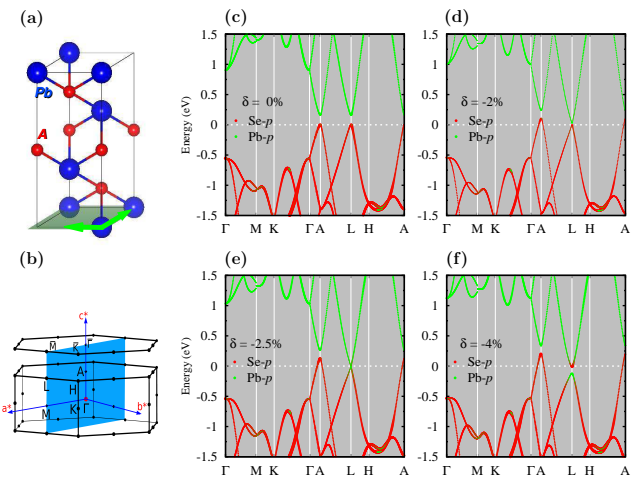}
\caption{(a) Chosen unit cell for (111)-oriented films
with highlighted in-plane biaxial strain directions and (b) corresponding Brillouin zone. (c)-(f) Band structure evolution of PbTe as a function of compressive
strain with projected Pb and Se $p$ states, showing first an insulator to semi-metal transition and then a band inversion at
the $L$ point.
}\label{fig2}
\end{figure}

{\it (111)-oriented systems.}
In addition to the volume and anion-cation distances modulation, 
a rhombohedral distortion of the unit cell,
characterized by the angle $\alpha\neq60^\circ$, is induced by a biaxial strain along the (111) plane.  The rhombohedral angle
evolves as
$\cos\alpha=[3-2\delta(4D^{111}+1)+\delta^2(4{D^{111}}^2-1)]/[6+4\delta(1-2D^{111})+2\delta^2(1+2{D^{111}}^2)]$, and gets
smaller/larger than 60$^\circ$ at compressive/tensile strain.
The main consequence of this rhombohedral distortion is to make the $A$ and $L$ points ($Z$ and $L$ in rhombohedral setting)
inequivalent, thus allowing, in principle, for topological transitions of $Z_2$ or TCI order if a band inversion occurs
at $A$ or $L$ rather than at $A$ and $L$ points, respectively.
When looking at the strained electronic structures of (111)-oriented compounds, the energy differences between VBM abd CBM
at $A$ and $L$ decrease (increase) under compressive (tensile) strain for all lead-salt chalcogenides.
However, hole and electron pockets appear before any band inversion takes place, as shown in Fig. \ref{fig2}.
Therefore, the topological phase-transitions never occur, being hindered by a transition to a semi-metal
phase . 

{\it Surface states.}
Due to the presence of (110) mirror planes, the band inversion at $R$ points in the strained (001)-compounds implies
 a non-zero mirror Chern number and the existence of two pairs of  spin-polarized surface
states with opposite mirror eigenvalues along the lines $\bar{X}-\bar{\Gamma}-\bar{X}$, crossing at four Dirac points off
the high-symmetry points $\bar{X}$.
\cite{Hsieh_ncomm, Fu_111,Safaei}
This is a consequence of the unique electronic topology of topological crystal insulators protected by mirror symmetries,
and its realization in rock-salt chalcogenides is due to the projection of two $R$ ($L$) points onto the same momentum on
the surface Brillouin zone; the resulting interaction between $L$-valleys is the main responsible for the displacement of
the Dirac cones away from the high-symmetry points $\bar{X}$, while a surface-state hybridization (which is
symmetry-prohibited along the mirror plane) forbids any band-crossing but along the $\bar{X}-\bar{\Gamma}$
line. The location $\Lambda$ of Dirac points away from $\bar{X}$ has been recently deduced in SnTe by
introducing a $k\cdot  p$ model including inter-valley scattering up to linear terms in $k$\cite{Fu_111}.
Within this approach,
$\Lambda=\pm \sqrt{m^2+d^2}/v_D$ along the high-symmetry line $\bar{X}-\bar{\Gamma}$, where $m,d$ describe the
energies of two Dirac cones centred at $\bar{X}$ and their hybridization, respectively, while $v_D$ is the Dirac velocity
at $\bar{X}$. The quantity $\mathcal{E}_0=2\sqrt{m^2+d^2}$ also describes the energy separation between surface conductance and
valence bands at $\bar{X}$, and as such it strongly correlates to the bulk gap.\cite{Tanaka_prb,Fu_111} Therefore, the
strain control of the bulk gap, as shown in Fig. \ref{fig1}, should in principle
also allow for a tunability of the Dirac cones location. 

\begin{figure}
\includegraphics[width=\columnwidth]{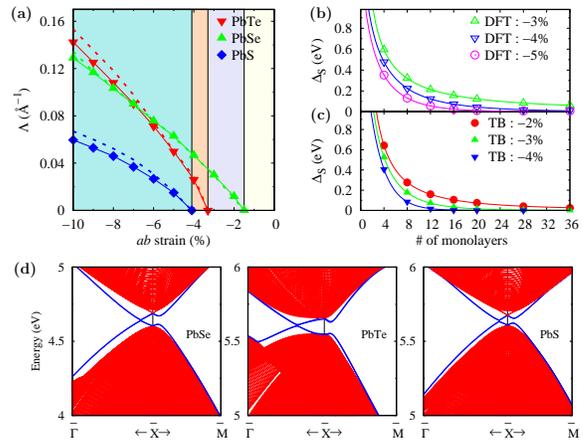}
\caption{{ (a) Tunability of the location $\Lambda$ of surface-state cones in strained Pb$A$; dashed lines represent the estimated
$\Lambda=\sqrt{m^2+d^2}/v_D$ in the framework of $k\cdot p$ model.} (b)-(c) Evolution of surface gaps as a
function of the slab thickness at selected values of strain in PbSe, as evaluated via  $nn$ TB or DFT-derived TB
parametrizations. (d) Zoom of {\it ab initio} surface states
for PbSe, PbTe at
a compressive strain $\delta=-4\%$ and PbS at $\delta=-5\%$, clearly displaying Dirac cones offcentered from the high-symmetry point $\bar{X}$.
}\label{fig3}
\end{figure}

The calculated surface states, as shown in
Fig. \ref{fig3}(d), that appear concomitantly with  the band-inversion, confirm this general picture. Within the
TB approach we could also study the evolution with strain of $\Lambda$, and estimate the $k\cdot p$ model parameters $m,d$
and $v_D$. As expected, $\mathcal{E}_0$ is found to
increase monotonically after the topological phase transitions in all Pb$A$; as a consequence, $\Lambda$ moves farther and
farther
away from $\bar{X}$ (see Fig. \ref{fig3}(a) --- dashed lines are obtained via the analytical expression of $\Lambda$\cite{Fu_111}).
The different slopes found for different lead-salts can be ascribed to their
different Dirac velocities $v_D$, which in turn depend on material-dependent microscopic properties
 and are also affected by strain. It is also worth to notice that the simple formulas given in Ref. \onlinecite{Fu_111} were obtained at the lowest
level of approximation in $k$, and may not be accurate when the hybridization between Dirac cones becomes relevant at $k-$points
far from $\bar{X}$; this explains the observed discrepancy at large values of the strain as reported in Fig. \ref{fig3}(a).

Another interesting issue which can be experimentally relevant is the critical thickness required by a TCI film
to show robust Dirac cones at its surface. In fact, if the decay length $\xi$ of the protected surface states is larger than the
film thickness, bulk properties are not completely recovered in the film and the interaction between the top and bottom
surface states is expected to open a hybridization gap $\Delta_S$.
In order to estimate the effect of this interaction, we have studied
the evolution of $\Delta_S$  as a function of the slab thicknesses, as shown in Fig. \ref{fig3}(b)-(c) for PbSe.
The decay length appears to be affected by the strain; in
particular, the critical thickness is found to get extremely large very close to the topological-phase transition
point,
rapidly diminishing as the bulk gap is increased. We notice that the $nn$ TB parametrization may well
be too crude for a realistic estimate of the decay length; in fact, we found that realistic TB parameters as obtained via
projecting DFT bands onto Wannier functions systematically result in larger decay lengths. On the other hand, the strain dependence of
$\xi$ is qualitatively correct, and the critical thickness drop off from $\sim \unit[50]{nm}$ to $\lesssim \unit[10]{nm}$
as increasing the compressive strain from $-3\%$ to $-5\%$ in PbSe [Fig. \ref{fig3}(c)].

In conclusion, we predict that compressive strains in the range between $-2\%$ and $-5\%$ turn all lead-salt
chalcogenides in TCI when grown on top of (001)-oriented substrates. Our prediction are based on both relativistic
$nn$ TB approach, as well as on accurate DFT calculations via HSE hybrid functionals, that are known to 
be required for a sufficient accuracy
especially for PbTe\cite{Kresse}; this explains the smaller critical strain previously predicted for strained PbTe from LDA
calculations\cite{huang_matexp}
as compared to our predicted value $\delta_c\lesssim -3.3$.
 As a consequence of the unique feature of 
mirror-symmetry protected TCIs, namely the interaction through inter-valley scattering of two Dirac cones,
strain tuning of the bulk gap results in a
potential control of topological properties. In particular surface-state cones may be moved at
different $k$ points as a function of strain, thus paving the way to new potential
applications requiring, e.g., Fermi-surface matching.
Such tunability of the $\bm k$-space location of Dirac cones has been already observed in
Pb$_{1-x}$Sn$_x$Te as a function of doping\cite{Tanaka_prb}; our findings further suggest to exploit this unique
feature of TCI  in strained-induced topological lead chalcogenides and/or in strained SnTe. We also predict that for
suitably chosen degree of strain, conducting surface states should appear in epitaxial films as thin as 
$\sim\unit[10]{nm}$. 
Our theoretical predictions naturally call for an experimental verification, as these conducting surface states should be
directly accessible to e.g. spin-resolved ARPES measurements.

\begin{acknowledgments}
We acknowledge PRACE for awarding us access to resource MareNostrum based
in Spain at Barcelona Supercomputing Center (BSC-CNS).
\end{acknowledgments}


\end{document}